\documentclass[multphys,vecphys]{svmult}

\usepackage{makeidx}     
\usepackage{graphicx}    
\usepackage{multicol}    

\makeindex             

\begin{document}

\title*{Simulated Radio Images and Light Curves of SN 1993J}
\author{Vikram V.~Dwarkadas\inst{1}\and
Amy J.~Mioduszewski\inst{2}\and Lewis T.~Ball\inst{3}}
\institute{ASCI FLASH Center, Univ of Chicago, 5640 S.~Ellis Ave,
Chicago IL 60637 \texttt{vikram@flash.uchicago.edu} \and NRAO,
P.O. Box O, Socorro, NM 87801 \texttt{amiodusz@nrao.edu} \and CSIRO
ATNF Parkes Observatory, P.O. Box 276, Parkes NSW 2870, Australia
\texttt{lewis.ball@csiro.au} }
%
%
\maketitle 

\abstract We present calculations of the radio images and light curves
from supernovae, based on high-resolution numerical simulations of the
hydrodynamics and radiation transfer in a spherically symmetric
medium. As a specific example we model the emission from SN1993J. This
supernova does not appear to be expanding in a self-similar fashion,
and cannot be adequately fitted with the often-used analytic
mini-shell model. We present a good fit to the radio evolution at a
single frequency.  Both free-free absorption and synchrotron
self-absorption are needed to fit the light curve at early times, and
a circumstellar density profile of $\rho \sim r ^{-1.7}$ provides the
best fit to the later data. Comparisons of VLBI images of SN1993J with
synthetic model images suggest that internal free-free absorption
completely obscures emission at 8.4~GHz passing through the center of
the supernova for the first few tens of years after explosion.

\section{Introduction}
\label{sec:1}

Radio Supernova (RSN) light curves are characterized by an initial
rapid increase of radio flux to maximum, followed by a power-law
decrease with time. High brightness temperature indicates a
non-thermal origin for the emission, and it is now accepted that the
emission is synchrotron in origin, due to the spiraling of
relativistic electrons in a magnetic field. The wavelength-dependent
turn-on suggests that the initial rise in the light curve is due to
decreasing absorption as the shock expands in an optically thick
medium.

In the past, the computation of SN radio light curves has been
accomplished mainly via semi-empirical methods, primarily the
mini-shell model developed by Chevalier (\cite{RAC82}) and extended by
Weiler et al.(\cite{WSP96}). These empirical methods assume a
self-similar evolution for the SN shock front, with the shock radius
expanding as a power law in time r$_{s} \propto t^m$, where $m$ is a
constant (\cite{RAC82}). The energy distribution of the relativistic
electrons is assumed to be a power law of constant spectral index, and
the energy density of the relativistic electrons and that of the
magnetic field are assumed to each be a constant fraction of the
thermal energy density behind the expanding supernova shock.

In this article we outline a more robust and general technique for
computing the radio light curves of young SNe. Details of our method
are outlined in Mioduszewski et al.~(\cite{MDB01}). Herein we briefly
describe the technique and the major results for the well-studied SN
1993J, as well as update the results with comparisons to more recent
data.

\section{Methods and Techniques}

$\bullet$ We start with a computation of the hydrodynamic evolution of
the SN remnant, calculated using VH-1, a 3-dimensional,
finite-difference, high resolution, shock-capturing code based on the
Piecewise Parabolic Method. This step immediately distinguishes our
method from the mini-shell model, since we do not need to assume a
self-similar expansion for the SN shock. Our technique is applied to
SN 1993J, where the observations indicate that the expansion is not
self-similar.
 
$\bullet$ The light curves are produced by calculating the transfer of
radiation along a line-of-sight through the SN.

$\bullet$ We assume that the injection of relativistic particles at
the shock follows a power-law, $N(E) \propto E^{- \gamma}$.

$\bullet$ We start with spherically symmetric simulations. These give
pressure, density and temperature at each grid point and every
timestep.

$\bullet$ The code takes the simulations and rotates them to form a 3D
sphere embedded in a Cartesian grid.

$\bullet$ The synchrotron emission and absorption is computed along
each ray. If $N(E) = K E^{- \gamma}$, then emissivity $j$ and opacity
$\kappa$ are given by

$$ j \sim K\; B^{(\gamma +1)/2}\; {\nu}^{-(\gamma - 1)/2.} $$
$$ \kappa \sim K\; B^{(\gamma + 2)/2}\; {\nu}^{-(\gamma + 4)/2.} $$

$\bullet$ In this work we assume that both magnetic energy density
$u_B$ and relativistic particle energy density $u_{rel}$ are
proportional to the thermal energy density.

$$ u_B = {\xi}_{B}\; u_{th}, \hspace{1.5truein} u_{rel} =
{\xi}_{rel}\; u_{th} $$

$ \bullet $ We take into account the external free-free absorption,
which depends on the temperature and density profile of the ambient
medium.

$\bullet$ The emissivity and absorption are used to calculate the
optical depth $\tau$, and finally the intensity $I_{\nu}
(r)$. Integration is carried out using the trapezoidal rule.

$\bullet$ The result is a 2D array of surface brightness, which is
used to make an image of the source and calculate the total flux.

\section{SN1993J}

As a first example we focus on SN 1993J - one of the brightest and
best studied SNe in the Northern Hemisphere. The radio flux evolution
of SN 1993J has been followed in detail by Bartel et al.~(\cite{B00};
\cite{B02}) and Perez-Torres et al.,~(\cite{PAM01}, \cite{PAM02})
using VLBI.

If shock radius $R_s \propto t^m$, then $m$ is called expansion
parameter. For power-law models, $m$ is constant with time. But for SN
1993J, $m$ decreases with time (\cite{B00}; \cite{B02}). Therefore the
evolution is NOT self-similar, which perhaps implies that power-law
density ejecta are not a reasonable assumption. For the ejecta
structure we have therefore used Model 4H47 of Suzuki \& Nomoto,
(\cite{SN95}). In this case the ejecta density shows considerably more
structure than the commonly used ``power-law'' models.

In order to fit the X-ray and optical light curves, Nomoto et al.~find
that the CSM density profile must decrease more slowly than $r^{-2}$,
which would be the case for a steady wind. Fransson and Bjornsson
(\cite{FB98}), however find that an $r^{-2}$ profile is adequate to
fit the radio light curve. Therefore we have tried 3 CSM density
profiles, with density decreasing as $r^{-1.7}$, $r^{-1.5}$ and
$r^{-2.}$.

Free Free absorption (FFA) alone results in an exponential rise of the
light curve while synchrotron self absorption (SSA) alone results in a
power-law increase (Fig 1a). In order to fit the light curve we find
that both FFA and SSA must be included, and that a CSM density profile
of $r^{-1.7}$ provides the best fit to the observed light curve at
8.4~GHz (Fig 1b). The data is represented by dots in the figure.

\begin{center}
\resizebox{100mm}{!}{{
 \includegraphics{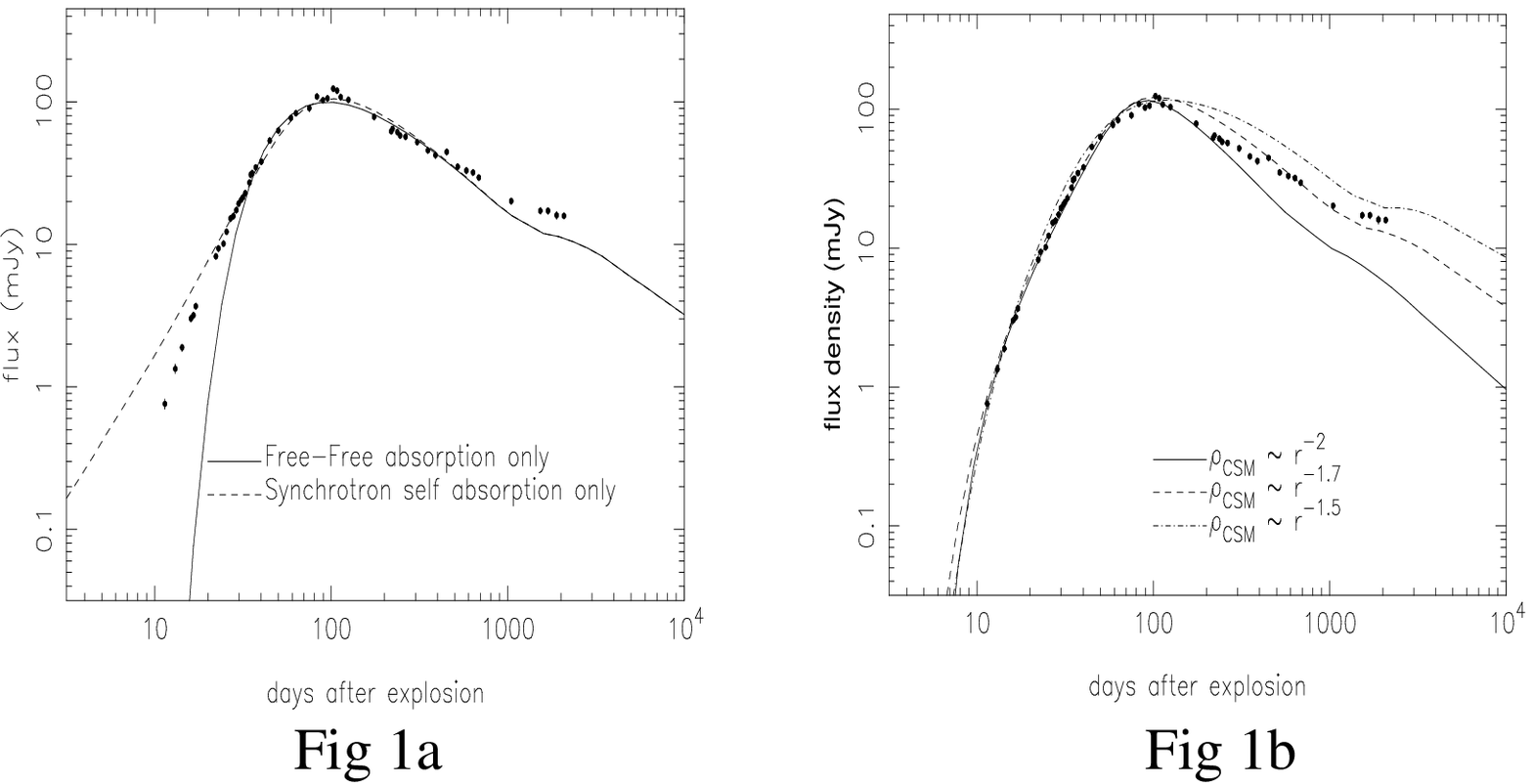}}}
\end{center}

As mentioned above, the expansion parameter $m$ (where R$_s \propto
t^m$) is not constant for SN 1993J, implying a non-self-similar
evolution. Fig 2a shows the evolution of the expansion parameter with
time from our simulation of the interaction of a SN with an ejecta
density profile described by the 4H47 model evolving in a medium whose
density varies with radius as r$^{-1.7}$.  The dashed lines in the
figure are error bars for the expansion parameter measured by Bartel
et al.~(\cite{B02}). Our simulations are broadly consistent with these
observations.

The model 4H47 exhibits local density maxima in the ejecta density
profile.  Impact of the reverse shock with one of these local density
maxima results in a sudden increase in the expansion parameter around
day 2300 (see Fig 2a). Observations by Bartel et al. (\cite{B02}) show
a similar rise in the expansion parameter (Fig 2b), albeit somewhat
earlier. It is possible that the change in the observed deceleration
is due to a change in the CSM density or some other cause unrelated to
the ejecta structure, but the coincidence is striking nevertheless. It
is also possible that in a more realistic multidimensional simulation
any local density maxima would be unstable and would quickly smooth
out.

\begin{center}
\resizebox{100mm}{!}{{
 \includegraphics{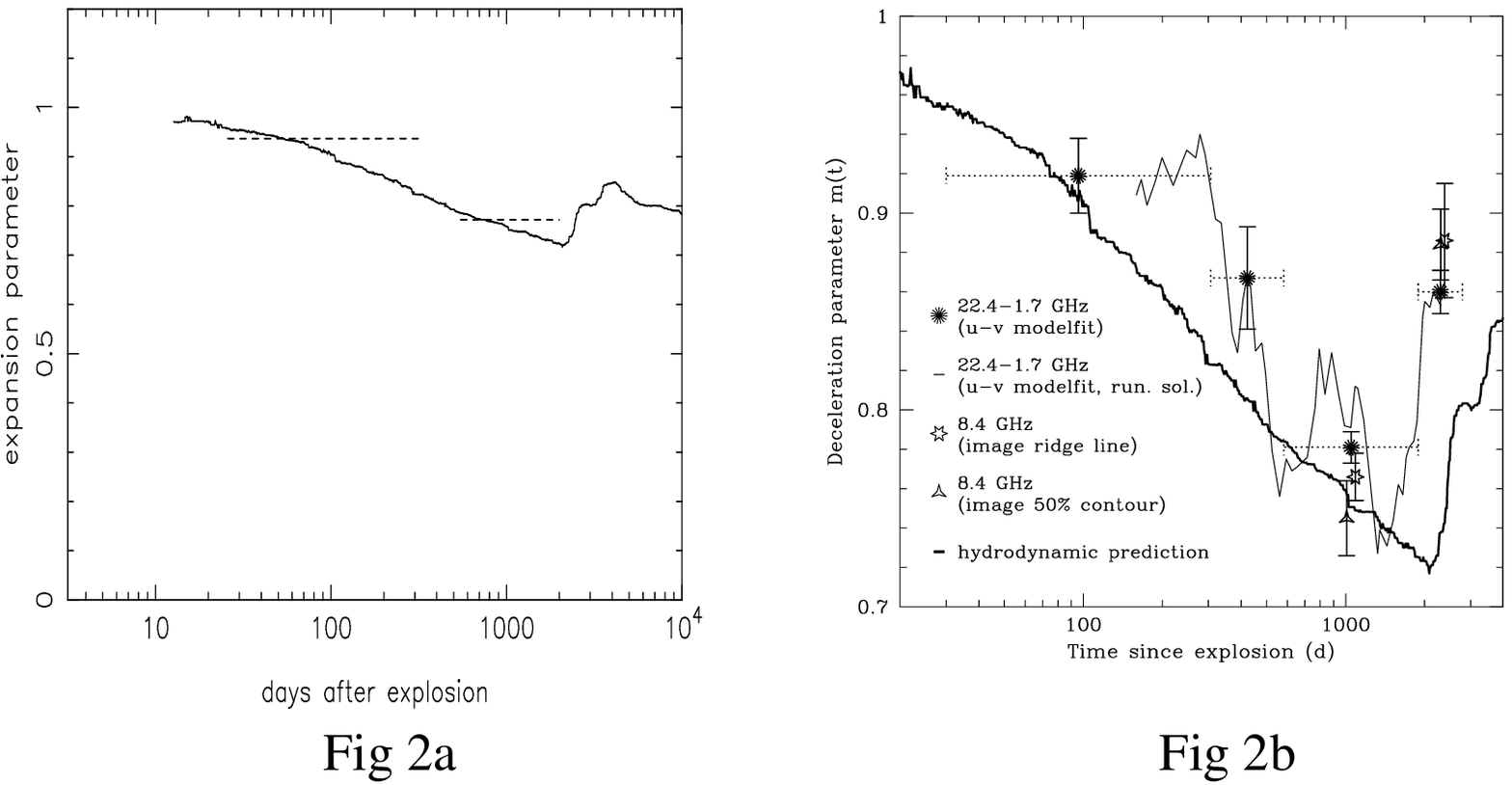}}}
\end{center}

\subsection{SN1993J Light Curves}

Our technique is to fit the observed light curve at one fiducial
frequency 8.4~GHz. The frequency dependence at all epochs is simply
$\nu^{-\alpha}$ (where $\alpha=0.55$ is the spectral index determined
by this best fit model of the light curve), and so the light curve at
any other frequency has the same shape as that at the fiducial
frequency. The observed flux densities from SN1993J and the
corresponding model light curves are shown in Figure 3. The different
frequencies indicated are: 22.5 GHz (filled boxes, dot-dashed line),
15.0 GHz (open circles, dotted line), 8.4 GHz (filled circles, solid
line), and 4.9 GHz (open triangles, dashed line)

\setcounter{figure}{2}
\begin{figure}
\centering
\includegraphics[height=5.5cm]{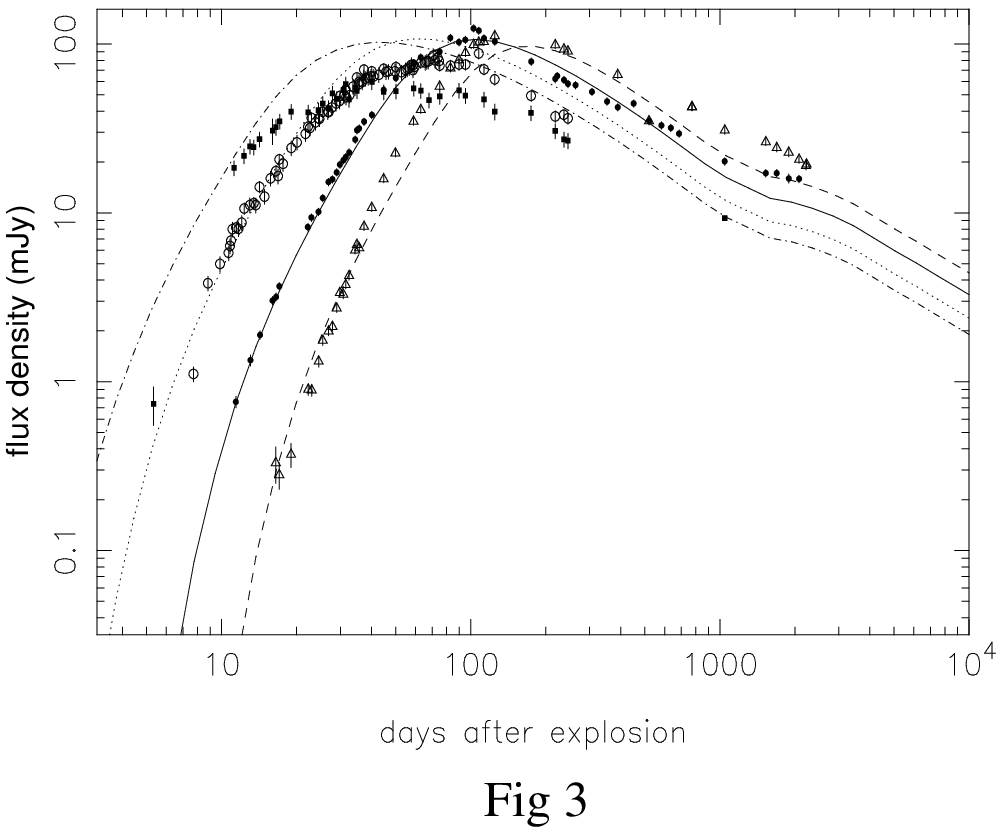}
\end{figure}

\subsection{Internal Absorption}

In the early stages, the internal free-absorption was found to be
important in order to match the surface brightness profile obtained
from our simulated images to that observed. Fig 4a shows the
azimuthally averaged profile of the observed 8.4 GHz emission from SN
1993J on day 1349 (solid line with error bars), compared to that
obtained from our simulations, excluding (dashed line) and including
(dotted line) internal free-free absorption, convolved with the same
beam as the image. The improvement in the fit when the internal
free-free absorption is taken into account is readily apparent. Fig 4b
shows the change in the light curve when the internal absorption is
taken into account. Our model suggests that the optical depth at 1~GHz
does not fall to unity until around 150 years after the
explosion. This suggests that a central radio pulsar would be
undetectable for many tens of years at frequencies below 1~GHz,
although multidimensional effects, especially instabilities, that have
not been included here may allow the radiation to escape somewhat
earlier.

\begin{figure}
\centering \includegraphics[height=5cm]{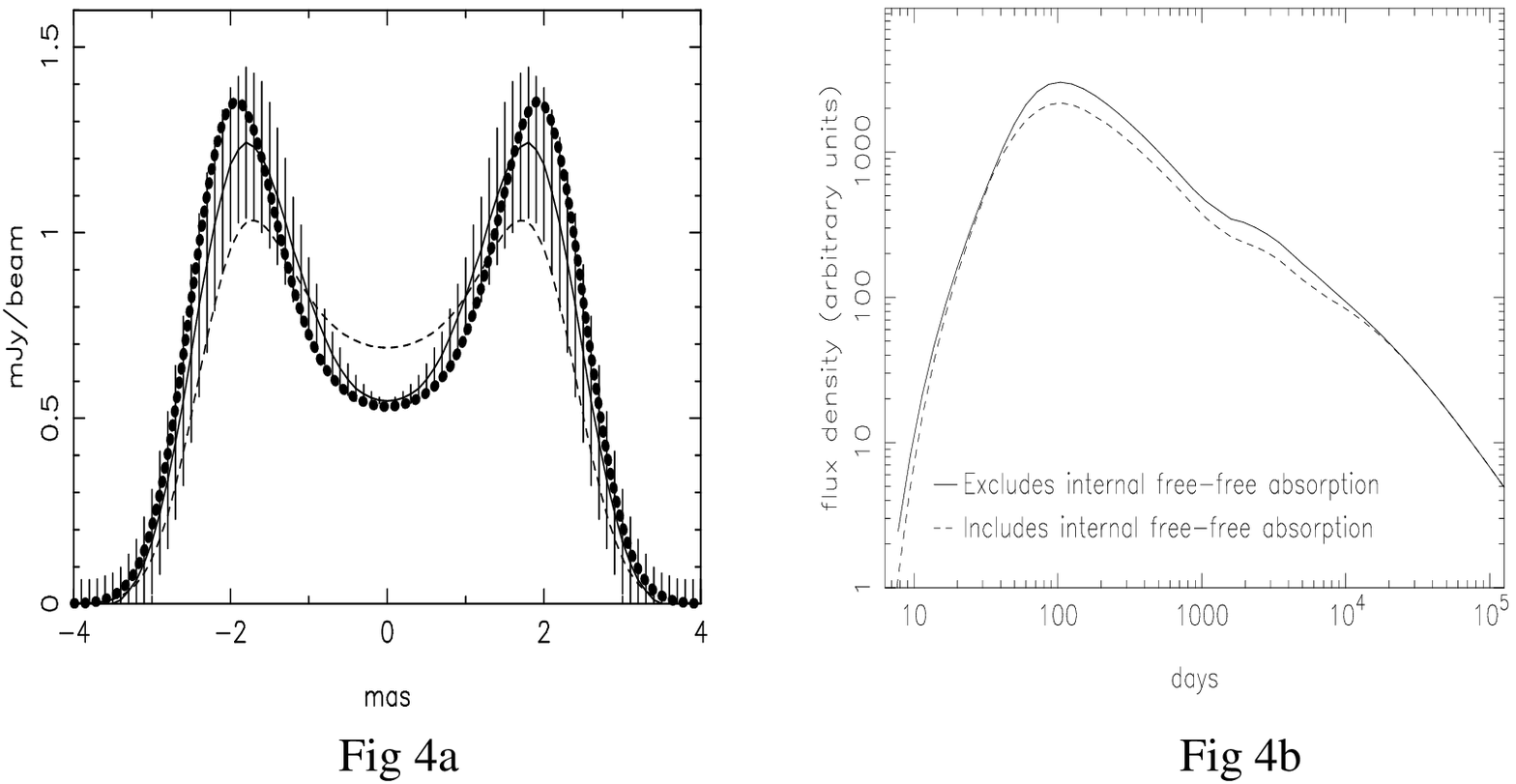}
\end{figure}

Further details can be obtained from Mioduszewski et
al.~(\cite{MDB01}).\\

{\bf{Acknowledgments}} Vikram Dwarkadas is supported by the
US.~Department of Energy grant number B341495 to the ASCI Flash Center
(U Chicago), and by Award \# AST-0319261 from the National Science
Foundation.

\printindex
\end{document}